\documentstyle[aps,graphicx]{revtex}
\def\bra{\langle} \def\ket{\rangle}
\def\<{\langle}
\def\>{\rangle}

\begin{document}
\title{Experimental realization of a fetching algorithm in  a 7 qubit NMR quantum computer
}
\author{Gui Lu Long$^{1,2,3,4}$
 and Li Xiao$^{1,2}$}
\address{$^1$ Department of Physics, Tsinghua University, Beijing 100084,
China\\
$^2$ Key Laboratory For Quantum Information and Measurements,
Beijing 100084,China\\
$^3$ Center for Atomic and Molecular Nanosciences, Tsinghua
University, Beijing 100084, China\\
$^4$ Institute of Theoretical Physics, Chinese Academy of
Sciences, Beijing 100080, China}
\date{\today}

 \maketitle
\begin{abstract}
Searching for marked items from an unsorted database is an
important scientific problem and  a benchmark for computing
devices as well. Using a 7-qubit liquid NMR quantum computer, we
have demonstrated successfully an hybrid quantum fetching
algorithm that finds marked items using only a single query. The
essential idea is the operation of quantum computers in parallel.
We gave the detailed pulse sequence  for coherent control of the
7 qubits. The pulse sequence demonstrated here is not only useful
for ensemble quantum computation, but also can be regarded as a
general purpose control-gate which is useful for experimental
design of quantum algorithms and general quantum information
processing task in other quantum computer schemes.  A
generalization of the algorithm that is scalable to arbitrary
qubit number is also provided.
\end{abstract}
\pacs{03.67.Lx, 76.60-k, 89.70.+c}

The perspective enormous power of  quantum computers(QCs) has
sparked intensive efforts worldwide. Various schemes have been
put forward. Liquid NMR\cite{cory97,gersh97,chuang98,knill98},
quantum dot\cite{qdot98}, cavity quantum
electrodynamics(QED)\cite{QED95}, linear ion trap\cite{trap94},
superconducting qubit\cite{supercon97}, solid-state
NMR\cite{kane98,yamamoto99,yamamoto02} are just a few examples in
this long list. By the way they measure, they can be put into two
main categories: single qubit-measurement quantum computer (SQC),
and ensemble qubit-measurement quantum computer(EQC). Typical
examples are Kane's proposed silicon quantum computer(SQC), and
the liquid NMR(EQC). SQC is the ideal system for quantum
computation. EQC is in fact an ensemble of many SQC copies. The
control and measurement in EQC, such as liquid NMR, is easier
than SQC for existing technology. Thus it advances farther in
demonstrating quantum information processing such as
Deutsch-Josza\cite{dj-1,dj-2,dj-3,dj-4}, quantum search and
counting\cite{gr-1,gr-2,gr-3,gr-4}, order-finding\cite{order},
quantum Fourier transformation\cite{zeng99,cory99}, Shor's
factorization\cite{vander01}, quantum error correction
code\cite{kni01}, dense coding\cite{fang00}, and GHZ-state
preparation \cite{kni00}.

Usually an ensemble is in a mixed state. Effective pure state(EPS)
technique is employed\cite{cory97,gersh97,chuang98,knill98} to
prepare the system so that the ensemble behaves like a pure
state. In liquid NMR, a measure that indicates to what extent the
system is close to a pure state is the polarization
$\epsilon$\cite{knill98}. If the polarization is equal to 1, the
ensemble  becomes a pure state ensemble, EQC is equivalent to
SQC. By increasing the magnetic field or employing polarization
transfer\cite{chuang01,luo98}, polarization can be increased. At
present, NMR-QC is operating at a regime with
$\epsilon\sim10^{-5}$. Liquid NMR QC may be used to investigate
the behaviour on representative systems and study problems that
will be encountered in future scalable technologies.

Redundancy in EQC not only makes qubit control and measurement
easy , but also provides source of additional computing power.
M\`{a}di, Br\"{u}schweiler and Ernst\cite{ma-br-er} put forward
Liouville space quantum computation(L-QC). In this mode, the
quantum parallelism and the classical parallelism in an ensemble
are exploited, and it can achieve exponential speedup for certain
problems which are impossible in a pure quantum computer.
Different from EPS where all the sub-ensembles are made into the
same state as much as possible, L-QC exploits the diversity of
states of constituent molecules in an ensemble. Notably, unsorted
database search can be exponentially fast in L-QC. Unsorted
database search is a benchmark problem for computing devices. It
is a mathematical NP-hard  problem and plays an important role in
computational complexity theory. It is widely used in science and
technology, for instance it can break DES-like encryption
schemes\cite{brassardscience}. To find a marked item from a
database with $N$-items, it requires about $N/2$ queries in a
classical computer. In a quantum computer, Grover's
algorithm\cite{grover} uses $O(\sqrt{N})$ steps, though a
significant square-root speedup but still exponentially slow.
When searching a database with unknown number of marked items, an
estimate of the number of marked states  is required because it
depends sensitively on this number. Furthermore it has been shown
that Grover's algorithm is optimal for quantum
computers\cite{ben97,boy98,zal99}. This optimality restriction
can only be broken off from outside of quantum computer and it
has been achieved by Br\"{u}schweiler in L-QC \cite{brusch}.
Br\"{u}schweiler's algorithm requires only $\log_2 N$ search
steps. For instance, to decipher a DES code, it is equivalent to
find a marked item from a  database with $N=2^{56}\simeq 10^{17}$
items. A quantum computer running Grover's algorithm requires
about $0.8 \times 2^{28}\simeq 2\times 10^{8}$ steps, whereas
Br\"{u}schweiler's algorithm requires only 56 search steps.
Recently, it is found that L-QC computing can achieve the
absolute maximum in unsorted database search, a single query
suffices to find every marked item using a hybrid quantum fetching
algorithm\cite{xiaolong} .

In this paper, we report the experimental implementation of this
hybrid quantum fetching algorithm in a NMR system with 7 qubits.
The implementation requires coherent quantum control of all the 7
qubits. Pulse sequences for the quantum gate operations are given
explicitly. We also scale up the algorithm to arbitrary qubit
number which can be used in scalable ensemble quantum computer.
In particular, we found that the algorithm can be implemented
directly from thermal equilibrium.

Here we briefly introduce the fetching algorithm, details can be
found in Ref.\cite{xiaolong}. In operator-product formalism , a
pure-state $|\phi_m\>$ is represented
 by direct product of the polarization
operators\cite{ma-br-er}, $|\phi_m\>=|\alpha \alpha \beta
...\alpha \beta\> \Longleftrightarrow \rho_m=I_1^{\alpha}
I_2^{\alpha} I_3^{\beta} ... I_{n-1}^{\alpha} I_n^{\beta}$ where
$|\alpha\>$, $|\beta\>$ are the spin up and down eigenstates of
$I_{z}$, the Pauli spin operator respectively. The one-qubit
state $|\alpha\>$ and $|\beta\>$ can be expressed in terms of the
polarization operator, namely $2I^{\alpha}=2|\alpha\>\<\alpha|=
(\textbf{1}+2I_z), \;\; 2I^{\beta}=2|\beta\>\<\beta|=
(\textbf{1}-2I_z)$. In L-QC with 7-qubit, one qubit is used as
ancilla as required by L-QC, the other six qubits represent a
database with $2^6=64$ items, namely $|0\ket=|000000\>$,
$|1\ket=|000001\>$, ...,$|63\ket=|111111\ket$. To start, the L-QC
system is prepared in a mixed state, apart from a scaling factor
\begin{eqnarray}
\rho_0=I^\alpha_0=|0,0\ket\bra0, 0|+|0,1\ket\bra0,
1|+...+|0,i\ket\bra0, i|+...+ |0,63\ket\bra 0,63|, \label{state0}
\end{eqnarray}
 where we have written the ancilla qubit state explicitly for clarity,
 namely $|0000000\ket=|0,0\ket$. Then the query function $f$ is
applied to the system, which changes state (\ref{state0})  into
\begin{eqnarray}
\rho_1=|f(0),0\ket\bra f(0),0|+...+ |f(i),i\ket\bra f(i),
i|+...+|f(63),63\ket\bra f(63),63|. \label{state1}\end{eqnarray}
 For constituent(to distinguish pure state from mixed state, we call a part in a
pure state as component, and a part in a mixed state as
constituent) corresponding basis $i$, query function does nothing
to it if $i$ does not satisfy the query, otherwise it flips the
ancilla qubit.  After the implementation of query, a read-out
pulse is applied to the ancilla qubit and its spectrum is
measured. Those constituents that satisfy the query have their
ancilla qubit flipped, and on the ancilla qubit spectrum, the
peaks corresponding to these constituents will have their peaks
inverted. The item that a peak corresponds is determined by the
peak frequency. Then by looking at the ancilla qubit spectrum, one
can easily read out the marked items.

The experiment is carried out on a VARIANT INOVA-600MHz
spectrometer. The liquid sample is $^{13}C$ labeled crotonic acid
with formula $C^1 H_3^3 C^2 H^1=C^3 H^2 C^4 O_2 H$, which is
manufactured by Cambridge Isotope Laboratories Inc. The same
substance has been used in Ref.\cite{kni00} to prepare GHZ state.
The chemical structure and the $J$ couplings can be found in
Fig.2 of Ref.\cite{kni00}. The seven qubits are the four carbon
nuclear spins, two proton spins in the middle connected with the
middle two carbon. The 3 methyl proton spins(denoted M) have the
same chemical shift, and they are treated as a single qubit. The
sample is firstly dissolved in a 0.5 ml acetone and then injected
into a 5 mm tube, deaired and sealed. The ancilla qubit is chosen
as $C^2$. Nuclear spins of $H^1$, $C^3$, $C^1$, $H^3$, $C^4$ and
$H^2$ are assigned as qubit 1, 2, 3, 4, 5 and 6 respectively, in
decreasing order by the $J$ magnitude of respective nuclear spin
with the ancilla bit. Our measured $J$ values agree very well
with that in Ref.\cite{kni00}. The $J_{0i}$ coupling between
qubit 0 and the rest six qubits are(in Hz): 156.0, 69.7, 41.6,
$-7.1$, 1.4, $-0.7$ for $i$ from 1 to 6 respectively. If $J_{0i}$
is positive such as qubit 1, 2, 3 and 5,  we assign
$|\alpha\ket=|0\ket$ and $|\beta\ket=|1\ket$, and if $J_{0i}$ is
negative such as qubit 4 and 6, we assign $|\alpha\ket=|1\ket$
and $|\beta\ket=|0\ket$. In doing so, the peaks in the ancilla
qubit spectrum will represent numbers monotonically in increasing
order starting from $|000000\ket$ with frequency
$\omega_0+\sum_{i=1}^6\pi |J_{0i}|$ on the far left to
$|111111\ket$ with frequency $\omega_0-\sum_{i=1}^6\pi |J_{0i}|$
on the far right.

\begin{figure}
\begin{center}
\includegraphics[width=12cm]{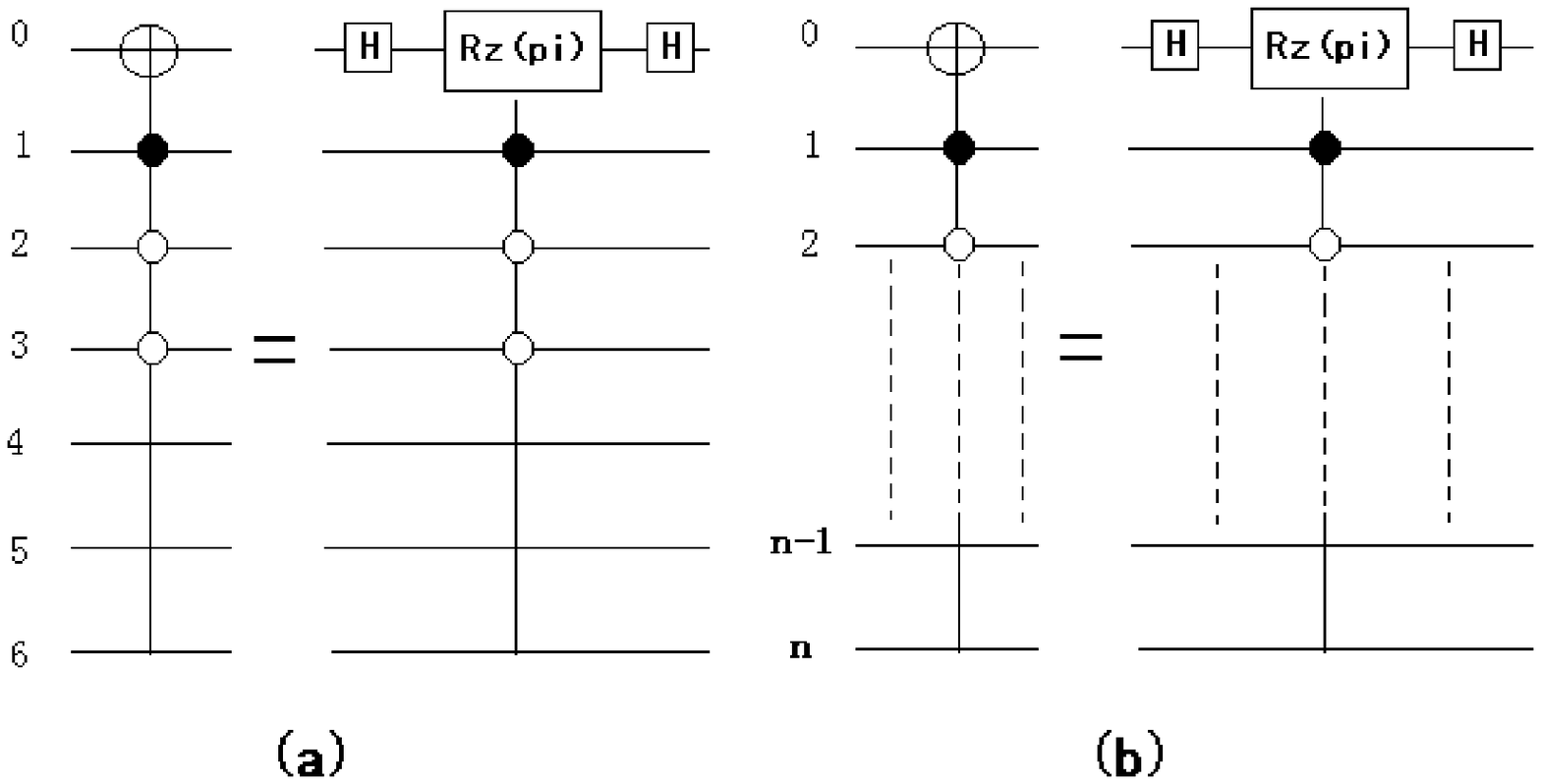}
\end{center}
\caption{Quantum network for query: (a)used in this work; (b) in
arbitrary qubit number system. $H$ is a Hadmard-like
transformation. A solid dot at the intersection with a qubit line
means 1, and circle means 0. $R_z(pi)$ is a $\pi$ rotation about
the z-axis.} \label{f2}
\end{figure}

\begin{figure}
\begin{center}
\includegraphics[width=10cm,angle=-90]{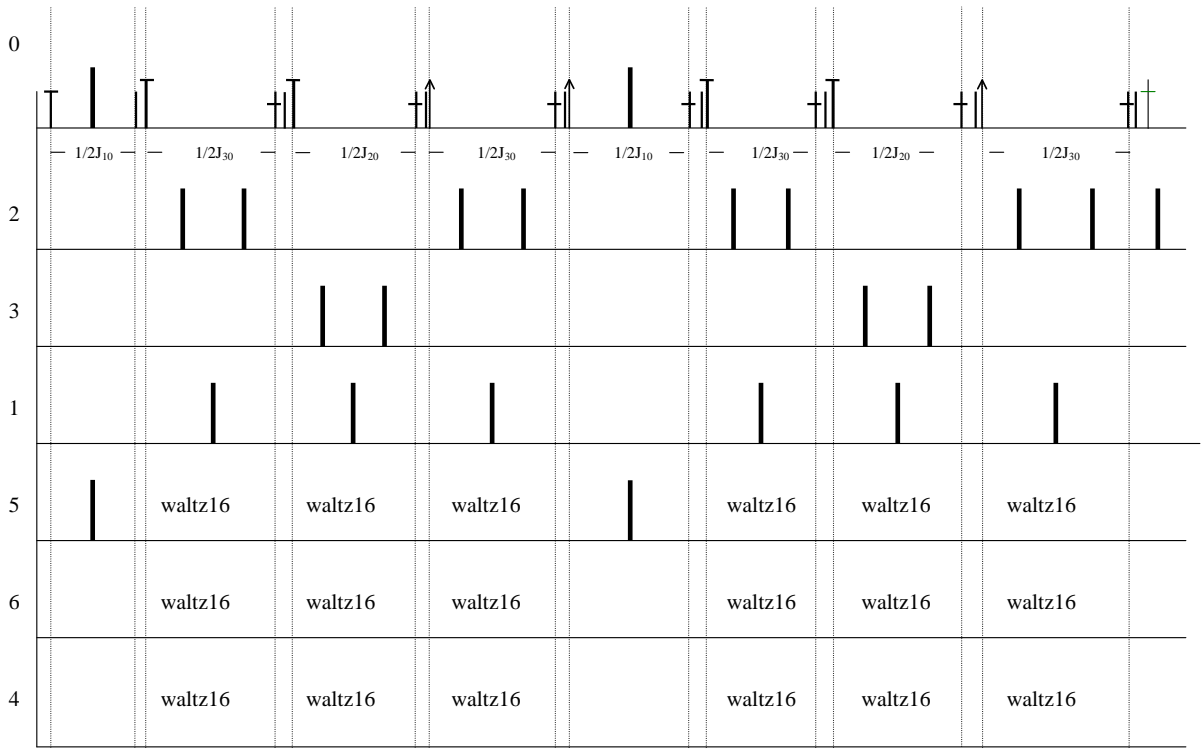}
\end{center}
\caption{The pulse sequence for the fetching algorithm. The
symbols(in brackets with sub-total number) for different pulses
(in degrees) are: 180 along $y$(thick long line, 23), 90 along
$y$(thin short line with a bar at top,1), 90 along $-y$(thin
short line with a bar at middle, 7), 112.5 along $y$(thin long
line with bar at top, 4), 22.5 along $-y$(thin long line with bar
at middle, 1), 67.5 along $y$(thin long line with arrow, 3) and
90 along $x$(thin short line, 8). }\label{f3}
\end{figure}

The query function $f$ is important in unsorted database search.
It  is treated as a
blackbox\cite{brusch,biron98,xiaolongyanyang}. The query blackbox
used in this work is given in Fig.\ref{f2}a. It checks whether an
items's first 3 bits values are 100. The H-gate in the figure is
a Hadmard-like gate with the expression
\begin{equation}
{\rm H}^0=\exp(-i\;\pi \;I_{x}^0)
\exp(-i\;\frac{\pi}{2}\;I_{y}^0)={\imath\over
\sqrt{2}}\left(\begin{array}{rr}
  1 & 1 \\
  1 & -1
\end{array}\right).
\end{equation}
It is realized using NMR pulse sequence $R_x^0 (\pi) R_y^0
(\frac{\pi}{2})$. The superscript(subscript) is the
qubit(direction) on which the pulse is applied. The triple-qubit
controlled phase rotation in the query is explicitly as
\begin{equation}
 R_z^{ccc}(\pi)=\exp(-i{\pi}I_z^0{(1-2I_z^1)\over 2}{(1+2I_z^2)\over 2}
 {(1+2I_z^3)\over 2}).\label{rccc}
\end{equation}
It is clear that when the first three qubits are in values 1, 0
and 0 respectively, the expression ${(1-2I_z^1)\over
2}{(1+2I_z^2)\over 2} {(1+2I_z^3)\over 2}$ gives an identity
operator, and the whole expression in (\ref{rccc}) becomes a
rotation about the $z$-axis through $\pi$. Otherwise, one or more
of the factors becomes zero, and Eq.(\ref{rccc}) reduces to an
identity operator. This implements the triple-qubit controlled
 phase rotation. Further it can be reduced to basic
one- and two-qubit gates by the recursive formula
\begin{equation}
\exp(-i\lambda 2^n I_{k_1 z}I_{k_2 z}...I_{k_n z}I_{k_{n+1}
z})=V_n \exp(-i\lambda 2^{n-1}I_{k_1
z}...I_{k_{n-1}z}I_{k_{n+1}z})V_n^{+},\;\;\;\;\;\;(n\ge 2)
\label{emiao1}
\end{equation}
given in Refs.\cite{miao1}, where
\begin{equation}
V_n=\exp(-i \frac{\pi}{2}I_{k_{n+1}x})\exp(-i \pi I_{k_n z
}I_{k_{n+1}z})\exp(i\frac{\pi}{2}I_{k_{n+1}x})\exp(-i\frac{\pi}{2}I_{k_{n+1}y}).
\label{emiao2}
\end{equation}
The whole pulse sequence is drawn in Fig.\ref{f3}. It involves the
coherent control of all the 7 qubits. It consists of 47
spin-selective radio frequency pulses, together with several free
evolution intervals. The decoupling is implemented using the
compound decoupling pulse sequence WALTZ16\cite{waltz16}. The
total duration of the pulse sequence is about 90 $ms$ which is
well within the system's 2$s$ decoherence time
($T_2$)\cite{kni00}.

\begin{figure}
\begin{center}
\includegraphics[width=12cm,angle=-90]{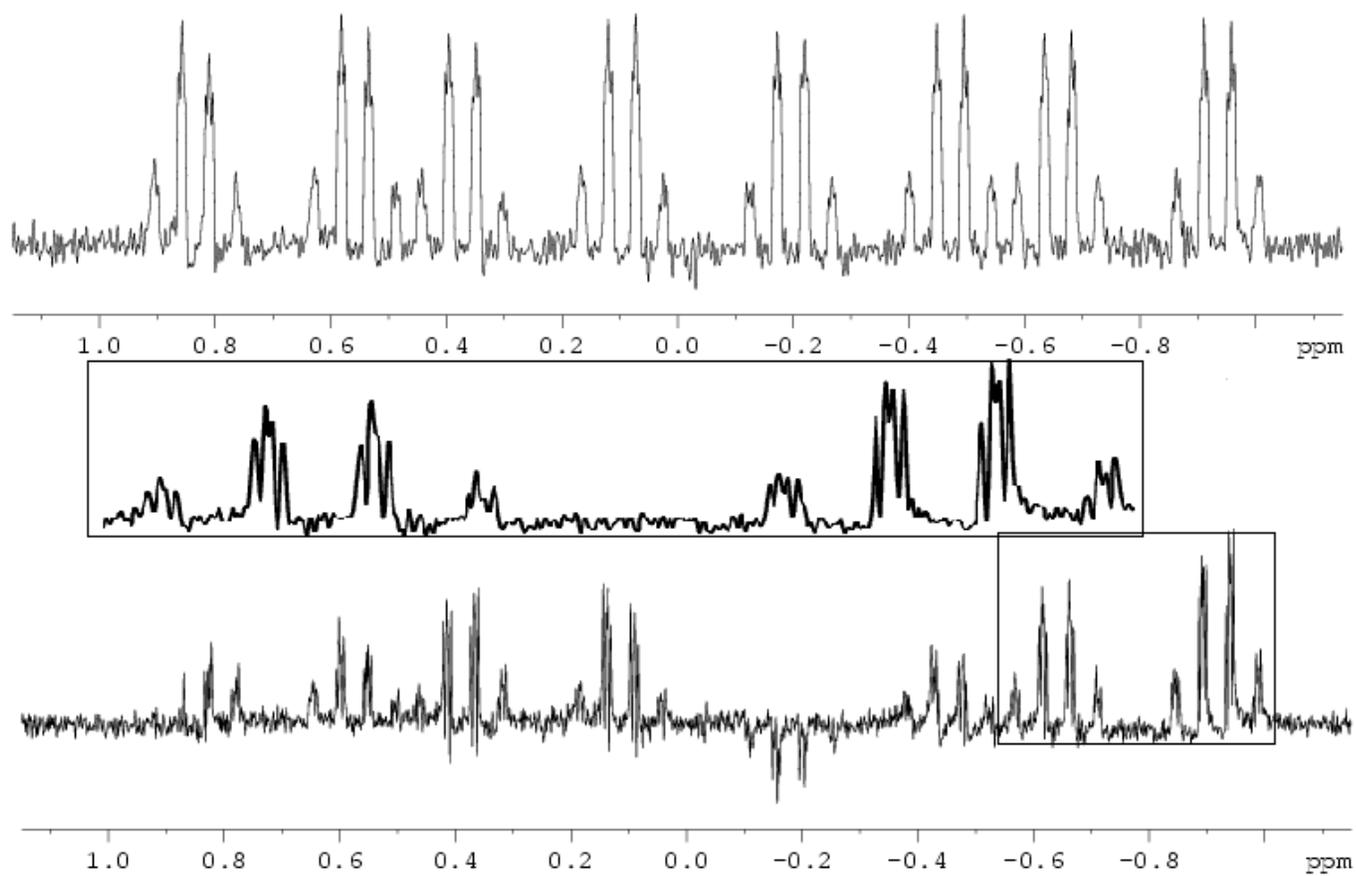}
\end{center}
\caption{The spectra before(top) and after(bottom) the
implementation of query.  The plot in the middle is a part of the
spectrum boxed at the bottom. Here it is seen that each peak in
the bottom spectrum has 4 sub-peaks. }\label{f4}
\end{figure}

We actually implemented the algorithm directly on the thermal
equilibrium state without preparing $I_0^{\alpha}$. In thermal
equilibrium, the density operator is
\begin{eqnarray}
\rho\simeq{1\over N}({\rm I}-\beta H)={1\over N}\left( {\rm
I}-\beta\sum_{i=0}^{n-1}\gamma_i B\sigma_z^i\right).
\label{thermal}
\end{eqnarray}
The state we need is $I^{\alpha}_0=({\rm
I}+\sigma_z^0)\sim\sigma_z^0$ and can be written as
$\rho'={1\over N}\left({\rm I}-\beta \gamma_0B\sigma_z^0\right)$,
apart from an identity operator and a scaling factor. $\gamma_i$
is the gyromagnetic ratio of qubit $i$, and $\beta=1/kT$. After
the implementation of query, only the ancilla quibit state is
changed,
\begin{eqnarray}
U_f\rho'U_f^{\dagger} &=&{1\over N}\left({\rm I}-\beta
U_f\gamma_0\sigma_z^0U_f^{\dagger}-\beta\sum_{i=1}^{n-1}\gamma_iB\sigma_z^i\right).
\label{ufstate}
\end{eqnarray}
Upon measurement, a selective read-out pulse is applied to the
ancilla qubit, the first and last terms in eq.(\ref{ufstate}) do
not contribute to the signal. The signal we acquire is only from
the second term which is the one we need. Thus we can perform the
algorithm directly from thermal equilibrium. Shown at the top of
Fig.\ref{f4} is the ancilla qubit spectrum at the thermal
equilibrium state. It represents the unsorted database. The
spectrum is divided into 8 major groups, and each group contains
4 peaks where two inner peaks are about 3 times as high as the
outer two short peaks. This is because qubit 4 is composed of 3
identical proton spins, and they produce 4 energy levels and thus
make 4 modulations on the ancilla bit spectrum with modulation
frequency $3J_{04}$ (all three spins up), $J_{04}$ (two spins up
and one spin down, triply degenerate), $-J_{04}$(two spins down
and one spin up, triply degenerate), and $-4J_{04}$ (all three
spins down). Because the multiplicity of levels due to qubit 4,
the number of peaks in spectrum is doubled. Thus within each
group the left two peaks(one short and one tall) corresponds to
qubit 4 at state 0, and the right two peaks corresponds to qubit
4 at state 1. While the doubling of levels gives additional work
in preparing an effective pure state in EPS QC, it is
advantageous for L-QC because this peak-doubling makes the
identification of peak inversion a lot easier. Within each peak in
Fig.\ref{f4}, there are still 4 finer peaks due to the different
states of qubit 5 and 6. In finer frequency resolution, they are
clearly separated and visible.  All the 128 peaks in the spectrum
correspond items from 0 at the far left and 63 at the far right.
It is interesting to notice the separation of the major groups.
Counting from left, the distance between group 2 and 3, 6 and 7
are shorter than that between groups 1 and 2, 3 and 4, 5 and 6, 7
and 8. The distance between 4 and 5 is the largest. These
distance differences can be easily understood from the $J$
parameter magnitude. The centroid positions of these peaks are of
the form $\pm |J_{01}|\pm |J_{02}|\pm |J_{03}|$.

Shown in bottom of Fig.\ref{f4} is the ancilla bit spectrum after
the implementation of the query. The spectrum is a result with
only 8 scans.  Shown in the middle of Fig.\ref{f4} is part of the
spectrum after the query, here it is clearly shown that there are
4 peaks within each peak. Even with this small scan number, it is
clearly visible from the figure that the peaks in the fifth group
have  their peaks inverted. The doubling of peaks due to qubit 4
makes the identification of peak inversion easier, in particular
by looking at the inner tall peaks. They correspond to items from
100000 to 100111 which are the items that we are searching. As
compared with that before the query, the height of peak is
reduced. This is due to the decoherence that occurs during the
computation. The result is striking because no measures have
taken to compensate the decoherence. The fetching of the marked
states is successfully demonstrated.

It is worth pointing that the algorithm demonstrated here is in
fact the parallel operation of quantum computers. The capability
of doing quantum computing in parallel lies in the unitarity
nature of QC operations. A function $f$ in a QC is carried out by
a unitary transformation $U_f$ which corresponds to a sequence of
radio frequency pulses and free evolutions. This pulse sequence
evaluates the function value for $|00...0\ket$, and the same
pulse sequence also evaluates the function value of $|11...1\ket$
and any other inputs! When applied to a superposition of basis
states, it evaluates the function values of all the basis states!
When applied to an ensemble in mixed state, the pulse sequence
transforms each constituent sub-ensemble of basis states into
their corresponding function values. The pulse sequence is
ubiquitous to all molecules in the ensemble and acts as a
super-commander to all the basis states. Each basis state goes
through the transformation required by the specific function
irrespective of its surroundings: alone, or lined up hand by hand
with other basis states in a superposed pure state, or scattered
here and there in different sub-ensembles in a mixed state.

The algorithm can be scaled up to systems with arbitrary $n$ qubit
system. An example for the query network for $n$-qubit system is
shown in Fig.\ref{f2}b. The multiple-qubit($i_1i_2..i_n$)
controlled phase rotation can be written explicitly
\begin{equation} R_z^{c^n} (\pi)=\exp(-i
\frac{\pi}{2^{n}}I_z^0(1+s_1(-1)^{i_1}2
I_z^1)...(1+s_{n-1}(-1)^{i_{n-1}}2 I_z^{n-1})(1+s_n(-1)^{i_n}2
I_z^n)),
\end{equation}
where $s_i$ denotes the sign of $J_{0i}$. Using formulas
(\ref{emiao1}) and (\ref{emiao2}), it can be decomposed into
basic one- and two- qubit gate operations. They might be used in
other ensemble quantum computers, such as the proposed schemes in
Refs.\cite{yamamoto99,yamamoto02}.

To summarize, we have  implemented a hybrid quantum fetching
algorithm in a 7 qubit NMR quantum computer.  The algorithm is
essentially the parallel operation of quantum computers.  A
scale-up version of the algorithm is provided. Furthermore it is
demonstrated that the algorithm can be implemented in a thermal
equilibrium state directly.

The authors are grateful for help from Prof. X. Z. Zeng, M. L.
Liu, J. Luo for Help. This work is supported in part by China
National Science Foundation, the National Fundamental Research
Program, Contract No. 001CB309308 and the Hang-Tian Science
foundation.

\newpage

\newpage
\end{document}